\documentclass[aps,prd,groupedaddress,preprintnumbers,%
  showpacs,showkeys,floatfix,amssymb,amsfonts,twocolumn]{revtex4-1}  
\usepackage{hyperref}
\usepackage[usenames]{color}
\usepackage{todonotes}
\usepackage{bbold}
\usepackage{amsmath}
\usepackage{multirow}
\usepackage{graphicx}
\usepackage{xfrac}
\newcommand{\tr}{\operatorname{Tr}}

\begin{document}
\title{Strange nucleon electromagnetic form factors from lattice QCD}
\author{
  C.~Alexandrou$^{1,2}$,
  M.~Constantinou$^{3}$,
  K.~Hadjiyiannakou$^{2}$,
  K.~Jansen$^{4}$,
  C.~Kallidonis$^{2,5}$,
  G.~Koutsou$^{2}$, and
  A.~Vaquero Avil\'es-Casco$^6$
}
\affiliation{
$^1$Department of Physics, University of Cyprus,  P.O. Box 20537,  1678 Nicosia, Cyprus\\
  $^2$Computation-based Science and Technology Research Center,
  The Cyprus Institute, 20 Kavafi Str., Nicosia 2121, Cyprus \\
  $^3$Department of Physics, Temple University, 1925 N. 12th Street, Philadelphia, PA 19122-1801, USA\\
  $^4$NIC, DESY, Platanenallee 6,  D-15738 Zeuthen,  Germany\\
  $^5$Department of Physics and Astronomy, Stony Brook University, 100 Nicolls Road, Stony Brook, NY 11794, USA \\
  $^6$Department of Physics and Astronomy, University of Utah, Salt Lake City, UT 84112, USA\\
}

\begin{abstract}
  We evaluate  the strange nucleon electromagnetic form factors using an
  ensemble of gauge configurations generated with two degenerate maximally twisted mass clover-improved
  fermions with mass tuned to approximately reproduce the  physical pion mass. In addition, we present  results for the disconnected light quark 
  contributions to the nucleon electromagnetic form factors. Improved stochastic methods are employed leading to high-precision results. The momentum dependence of the disconnected contributions is fitted using the model-independent z-expansion. We
  extract the magnetic moment and the
  electric and magnetic radii of the proton and neutron by including both connected and disconnected contributions.
 We find that the disconnected light quark contributions to  both electric and magnetic form factors are  non-zero and at the few percent level  as compared to the connected. The strange form factors are also at the percent level but more noisy yielding statistical errors that are typically within one standard deviation from a zero value. 
 
\end{abstract}
\pacs{11.15.Ha, 12.38.Gc, 24.85.+p, 12.38.Aw, 12.38.-t}
\keywords{Nucleon structure, Electromagnetic form factors, Disconnected, Strangeness, Lattice QCD}
\maketitle

\section{Introduction}

The electromagnetic form factors of the nucleon are important
quantities encapsulating information about the distribution of
electric charge and magnetism inside the proton and neutron. Namely,
at zero momentum transfer, electromagnetic form factors yield the
electric charge and magnetic moment, while from the slope of the form
factors at zero momentum transfer one extracts the nucleon
radii. Obtaining the individual quark contributions is a major
theoretical and experimental challenge, which can
reveal insights on the partonic structure of the nucleon. In
particular, the strange quark contribution, which is subdominant
compared to the up and down quark contributions, is especially
challenging to measure. The interference between the  weak and electro-magnetic amplitudes leads
to a parity-violating asymmetry in the elastic scattering cross section for right- and left-handed electrons, which gives information on the strange form factors. Measuring the parity-violating electroweak asymmetry in elastic scattering of polarized electrons from protons, the HAPPEX collaboration~\cite{Aniol:2004hp} extracted the linear combination of strange form factors
$G^s_E + 0.392G^s_M = 0.014 \pm 0.020 \pm 0.010 $ at $Q^2=0.48$~GeV$^2$ which was found to be compatible with zero, where $G^s_E$ is strange electric and $G^s_M$ the strange magnetic proton form factor.  The A4 experiment at MAMI~\cite{Maas:2004ta}  finds a combination
$G^s_E + 0.225G^s_M = 0.039 \pm  0.034$ at $Q^2=0.23$~GeV$^2$, slightly non-zero within errorbars, while the SAMPLE experiment~\cite{Hasty:2001ep} determined the strange magnetic form factor $G^s_M (Q^2 = 0.1) = 0.14 \pm 0.29  \pm 0.31$ which is consistent with zero.
A combined analysis
of proton and neutron electromagnetic and weak form factors from
elastic electron-nucleon scattering mediated by photon and $Z^0$
exchange provides more recent estimates for the electric and magnetic form
factors (see Refs.~\cite{Ahmed:2011vp, Baunack:2009gy, Androic:2009aa}
for some recent experimental results). These studies also deliver results
consistent with zero for the strange quark contribution, and as such,
provide limits on the contribution of strange quarks in the
distribution of nucleon charge and magnetization.

Lattice QCD allows for a first principles calculation of the nucleon form factors. 
In lattice QCD, the calculation of the individual quark contributions to nucleon matrix elements requires the so called disconnected contributions, such as the one shown in Fig.~\ref{fig:thrp}. A number of lattice QCD calculations exist for the isovector form factors, or equivalently  the
combination $(G^p_{E,M}-G^n_{E,M})$, in which the disconnected contributions cancel in the isospin limit, as well as the isoscalar $(G^p_{E,M}+G^n_{E,M})$ 
combination neglecting disconnected contributions, of the electric and magnetic Sachs form
factors using simulations with
near-physical~\cite{Capitani:2015sba,Bhattacharya:2013ehc,Green:2014xba,Alexandrou:2017ypw} and higher than physical~\cite{Alexandrou:2006ru,Alexandrou:2011db,Alexandrou:2013joa} pion masses. Disconnected
contributions have only recently been calculated, typically using  larger than physical pion masses~\cite{Green:2015wqa}. In this study, we evaluate both the light and strange disconnected quark loops to
high-statistical precision using an
ensemble of two- degenerate twisted mass fermions with a clover term with quark mass tuned to yield a pion mass of about
$130$~MeV~\cite{Abdel-Rehim:2015pwa}. The disconnected quark loops are
estimated using improved stochastic techniques for several momenta and the nucleon two-point correlation functions
are computed using a number of final momenta,
allowing us to obtain the form factors from
multiple nucleon moving frames. We extract the magnetic moment, electric and magnetic radii by fitting the momentum dependence of the form factors to the model-independent  z-expansion~\cite{Hill:2010yb}. We use the connected contributions as calculated in Ref.~\cite{Alexandrou:2017ypw} to obtain results for the total quark contributions to the nucleon electromagnetic form factors, present new results for the strange quark contributions, and update the disconnected contributions for the light quarks.

The remainder of this paper is organized as follows: In
Section~\ref{Sec:LatEx}, we explain how we compute the nucleon matrix element within lattice QCD and 
in Section~\ref{Sec:Res} we provide the technical details of the calculation of the disconnected contributions, the analysis and results. In Section~\ref{Sec:Comp}, a comparison with other studies is performed and in Section~\ref{Sec:Concl} we summarize and
tabulate our findings.

\section{Lattice extraction} \label{Sec:LatEx}
The electromagnetic nucleon matrix element is decomposed in terms of two parity preserving form factors, the Dirac ($F_1$) and Pauli ($F_2$) form factors, given in Minkowski space by,
\begin{eqnarray}
  && \langle N(p',s') \vert j_\mu \vert N(p,s) \rangle = \sqrt{\frac{m_N^2}{E_N(\vec{p}\,') E_N(\vec{p})}} \times \nonumber \\
  && \bar{u}_N(p',s') \left[ \gamma_\mu F_1(q^2) + \frac{i \sigma_{\mu\nu} q^\nu}{2 m_N} F_2(q^2) \right] u_N(p,s).
  \label{Eq:Decomposition}
\end{eqnarray}
$N(p,s)$ is the nucleon state with initial (final) momentum $p$ ($p'$)
and spin $s$ ($s'$), with energy $E_N(\vec{p})$ ($E_N(\vec{p}\,') $)
and mass $m_N$. $q^2=q_\mu q^\mu$ is the momentum transfer squared where
$q_\mu=(p_\mu'-p_\mu)$ and $u_N$ is the nucleon spinor. The vector current $j_\mu$
is given by
\begin{equation}
  j_\mu(x) = j^l_\mu(x) + j^s_\mu(x)
  \label{Eq:VecCurr}
\end{equation}
with
\begin{equation}
  j^l_\mu(x) = e_u\, \bar{u}(x) \gamma_\mu u(x) + e_d\, \bar{d}(x) \gamma_\mu d(x),
\label{Eq:VecCurr_l}
\end{equation}
and
\begin{equation}
  j^s_\mu(x) = e_s\, \bar{s}(x) \gamma_\mu s(x),
  \label{Eq:VecCurr_s}
\end{equation}
where $(e_u,e_d,e_s)=(2/3,-1/3,-1/3)$ are the electric charges carried
by the up, down and strange quarks respectively. In this study, we use
the local vector current, therefore renormalization is necessary and has been computed 
non-perturbatively  using the  ${\rm RI^\prime_{MOM}}$
scheme~\cite{Gockeler:1998ye,Alexandrou:2012mt}.
Lattice artifacts have been evaluated in perturbation theory to 1-loop level and all orders in the lattice spacing and have been subtracted before taking the chiral and continuum limits~\cite{Alexandrou:2015sea}.

The nucleon matrix element on the lattice requires the evaluation of
three- and two-point correlation functions. The three-point function
in momentum space is given by
\begin{eqnarray}
 && C_\mu(\Gamma_\nu,\vec{q},\vec{p}\,';t_s,t_{\rm ins},t_0) = \sum_{\vec{x}_{\rm ins},\vec{x}_s} e^{i (\vec{x}_{\rm ins} - \vec{x}_0)  \cdot \vec{q}}  e^{-i(\vec{x}_s - \vec{x}_0)\cdot \vec{p}\,'} \times \nonumber \\
 && \tr \left[ \Gamma_\nu \langle J(t_s,\vec{x}_s) j_\mu(t_{\rm ins},\vec{x}_{\rm ins}) \bar{J}(t_0,\vec{x}_0) \rangle \right],
\end{eqnarray}
and the two-point function is given by
\begin{eqnarray}
  && C(\Gamma_0,\vec{p};t_s,t_0) = \sum_{\vec{x}_s} \tr \left[ \Gamma_0  \langle J(t_s,\vec{x}_s) \bar{J}(t_0,\vec{x}_0) \rangle \right] \times \nonumber \\
  && e^{-i (\vec{x}_s-\vec{x}_0) \cdot \vec{p}} \;,
\end{eqnarray}
where $J_N$ is the standard nucleon interpolating field:
\begin{equation}
  J_N(\vec{x},t)=\epsilon^{abc}u^a(x)[u^{b\intercal}(x)C\gamma_5d^c(x)],
\end{equation}
with $C=\gamma_0\gamma_2$ the charge conjugation matrix and $u$ and
$d$ are the up- and the down-quark fields respectively. $\Gamma_\nu$ is a
projector acting on spin space, with $\Gamma_0=\frac{1+\gamma_0}{4}$
projecting to unpolarized nucleons and
$\Gamma_k=i\gamma_5\gamma_k\Gamma_0$ projecting to nucleons polarized
in direction $k$.

\begin{figure}[ht!]
  \includegraphics[width=1\linewidth]{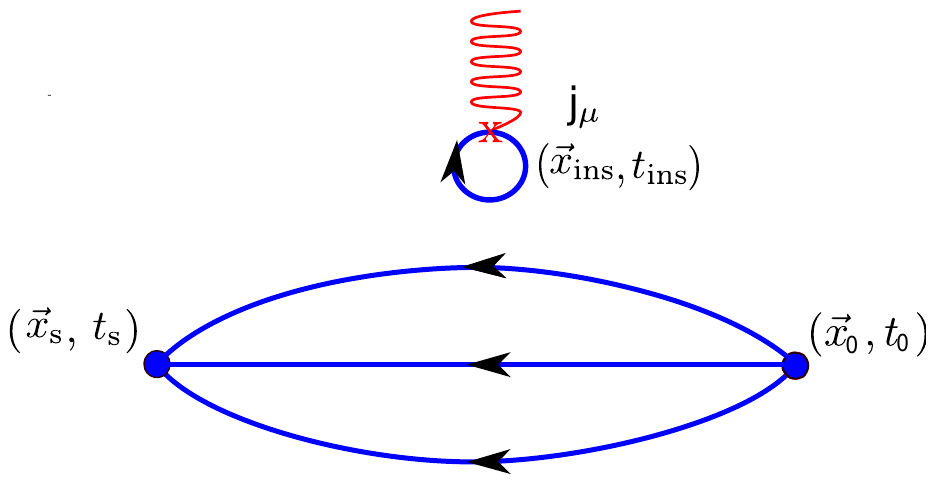}
  \caption{Disconnected three-point nucleon correlation function with source at
    $x_0$ and sink at $x_s$ with vector insertion $j_\mu$ at $x_{\rm ins}$. 
  }
  \label{fig:thrp}
\end{figure}

The three-point function receives contributions from both quark-
connected and disconnected terms. As mentioned, the connected contributions have been evaluated and
presented in Ref.~\cite{Alexandrou:2017ypw} for the same ensemble as the one used
here, as have preliminary results for the disconnected light quark contributions. In this work, we present a thorough analysis of the disconnected
contributions, depicted in Fig.~\ref{fig:thrp}, updating our results for light quarks
and showing results on the strange quark contributions not calculated previously.
 We use
Osterwalder-Seiler strange quarks~\cite{Osterwalder:1977pc} and tune the strange
quark mass to reproduce the experimental $\Omega^-$ mass. This
yields $a \mu_s=0.0259(3)$, where the lattice spacing $a=0.0938(3)$~fm as determined from the nucleon mass~\cite{Alexandrou:2017xwd},
yielding a renormalized strange quark mass at 2~GeV in the $\overline{\rm MS}$-scheme $m_s^R=108.6(2.2)$~MeV.

 To isolate the electromagnetic matrix element in the three-point
 function, an optimized combination of two-point functions is
 constructed to form the ratio,
\begin{eqnarray}
&&  R_\mu(\Gamma_\nu,\vec{p}\,',\vec{p};t_s,t_{\rm ins}) = \frac{C_\mu(\Gamma_\nu,\vec{p}\,',\vec{p};t_s,t_{\rm ins})}{C(\Gamma_0,\vec{p}\,';t_s)} \times \nonumber \\
&&  \sqrt{\frac{C(\Gamma_0,\vec{p};t_s-t_{\rm ins}) C(\Gamma_0,\vec{p}\,';t_{\rm ins}) C(\Gamma_0,\vec{p}\,';t_s)}{C(\Gamma_0,\vec{p}\,';t_s-t_{\rm ins}) C(\Gamma_0,\vec{p};t_{\rm ins}) C(\Gamma_0,\vec{p};t_s)}} \;\;.
\label{Eq:ratio}
\end{eqnarray}
In the large time limit, 
$R_\mu(\Gamma_\nu;\vec{p}\,',\vec{p};t_s;t_{\rm ins})\xrightarrow[t_{\rm ins}\rightarrow\infty]{t_s-t_{\rm ins}\rightarrow\infty}\Pi_\mu(\Gamma_\nu;\vec{p}\,',\vec{p})$
yielding a time independent plateau. Note that in Eq.~(\ref{Eq:ratio}),  $t_s$ and $t_{\rm ins}$ are relative to the source, $t_0$, which is omitted, and we will adopt this convention for the remainder of this paper.
When taking large time separations to obtain $\Pi_\mu(\Gamma_\nu;\vec{p}\,',\vec{p})$, one cannot set the source-sink time separation to arbitrarily large values 
since the noise-to-signal ratio grows exponentially. Therefore, one seeks  a window within  which the
source-sink separation is large enough for the excited states to be suppressed
while small enough to yield a good signal. We employ
Gaussian smearing~\cite{Gusken:1989qx,Alexandrou:1992ti} to increase
the overlap with the ground state and apply APE smearing~\cite{Albanese:1987ds} to the gauge links, with the same parameters
used in Ref.~\cite{Alexandrou:2017ypw}.

The Dirac and Pauli form factors, $F_1$ and $F_2$, are related to the electric Sachs ($G_E(Q^2)$) and magnetic
Sachs ($G_M(Q^2)$) form factors via:
\begin{eqnarray}
  && G_E(Q^2) = F_1(Q^2) - \frac{Q^2}{(2 m_N)^2} F_2(Q^2), \\
  && G_M(Q^2) = F_1(Q^2) + F_2(Q^2)
\end{eqnarray}
where $Q^2=-q^2$ is the Euclidean momentum transfer squared. The
combination of the projector $\Gamma_\nu$, the current insertion and
the initial and final momenta $\vec{p}$, $\vec{p}\,'$ leads to an
overconstrained set of equations relating
$\Pi_\mu(\Gamma_\nu;\vec{p}\,',\vec{p})$ to $G_E$ and $G_M$. We solve
by using the Singular Value Decomposition of the minimization problem
that arises. The expressions used are given in Appendix
~\ref{sec:appendix equations}. The same procedure has been followed
for extracting the axial and induced pseudo-scalar form factors in
Ref.~\cite{Alexandrou:2017hac}, where more details can be found. For
the results that follow, the analysis combines two values of the
final momentum, namely $\vec{p}\,'=\vec{0}$ and $\vec{p}\,'=\frac{2
  \pi}{L} \vec{\hat{n}}$.

In what follows  we use two analysis  methods to assess  excited
states contamination and extract the matrix element of the nucleon.\\
 \emph{Plateau method:} For specific $t_s$ one identifies a range of $t_{\rm ins}$ where the value of the ratio remains unchanged and performs a constant fit. This procedure is repeated for several $t_s$ seeking for convergence in the matrix element of the ground state.\\
\emph{Summation method:} Summing over $t_{\rm ins}$ in the ratio of Eq.~(\ref{Eq:ratio}) between the source and the sink gives,
\begin{equation}
  \sum_{t_{\rm ins}=a}^{t_s-a}  R_\mu(\Gamma_\nu,\vec{p}\,',\vec{p};t_s,t_{\rm ins}) = C + t_s \mathcal{M} + \mathcal{O}(e^{-\Delta E t_s})
\end{equation}
where $\Delta E$ is the energy gap between the ground state and the first excited state. The nucleon matrix element, $\mathcal{M}$, is extracted from the slope by fitting to a linear form. The summation method will be used to provide an estimate of the systematic error due to potential contamination from  excited states.

In Table~\ref{table:sim} we summarize the parameters of the simulation. Details on the determination of the nucleon and pion mass and the lattice spacing are given in Ref.~\cite{Alexandrou:2017xwd}.

\begin{table}[ht!]
  \caption{Simulation parameters. First row gives the $\beta$-value, the value of the clover parameter $c_{\rm SW}$, the lattice spacing and the  Sommer parameter $r_0$.}
  \label{table:sim}
  \begin{tabular}{c|r@{=}l}
    \hline\hline
    \multicolumn{3}{c}{$\beta$=2.1, $c_{\rm SW}$=1.57751, $a$=0.0938(3)~fm, $r_0/a$=5.32(5)} \\
    \hline
    \multirow{4}{*}{48$^3\times$96, $L$=4.5~fm} & $a\mu_l$  & 0.0009       \\
      & $m_\pi$      & 0.1304(4)~GeV\\
      & $m_\pi L$    & 2.98(1)      \\
    & $m_N$        & 0.932(4)~GeV \\
    & $m_N/m_\pi$  &  7.15(4)\\
      \hline\hline
  \end{tabular}
\end{table}

In Table~\ref{Tab:statistics} we tabulate the statistics used in this work. The disconnected quark loop entering the diagram of Fig.~\ref{fig:thrp} cannot be computed exactly, except
for very small lattices. In this work, we employ stochastic techniques combined with
the so-called \emph{one-end trick}~\cite{McNeile:2006bz} and
specifically its generalized version explained in detail in
Refs.~\cite{Alexandrou:2017hac,Alexandrou:2013wca,Abdel-Rehim:2013wlz} to  estimate  the disconnected quark loops.
The light quark loops are produced using  high-precision inversions employing deflation of the low modes to overcome  critical slowdown. For the computation of strange quark loops we employ the truncated solver method (TSM)~\cite{Bali:2009hu} to increase the statistics at low cost. Details for the tuning procedure followed  can be found in Ref.~\cite{Alexandrou:2017hac}. Note that we do not use any kind of dilution, therefore we invert each noise vector once.

\begin{table}[ht!]
\caption{The statistics of our calculation. $N_{\rm conf}$ is the number of gauge configurations analyzed and $N_{\rm src}$ is the number of source positions per configuration.  $N_r^{\rm HP}$ is the number of high-precision stochastic vectors used, and $N_r^{\rm LP}$ is the number of low-precision vectors used when employing the truncated solver method.} \begin{center}
\begin{tabular}{cccc|c}
\hline\hline
 Flavor & $N_{\rm conf}$ & $N_r^{\rm HP}$ & $N_r^{\rm LP}$ & $N_{\rm src}$ \\
\hline
light   & 2120 & 2250 &   -  & 100 \\
strange & 2057 & 63   & 1024 & 100 \\
\hline
\end{tabular}
\label{Tab:statistics}
\end{center}
\end{table}

\section{Analysis and results} \label{Sec:Res}
\begin{figure}[ht!]
  \includegraphics[width=\linewidth]{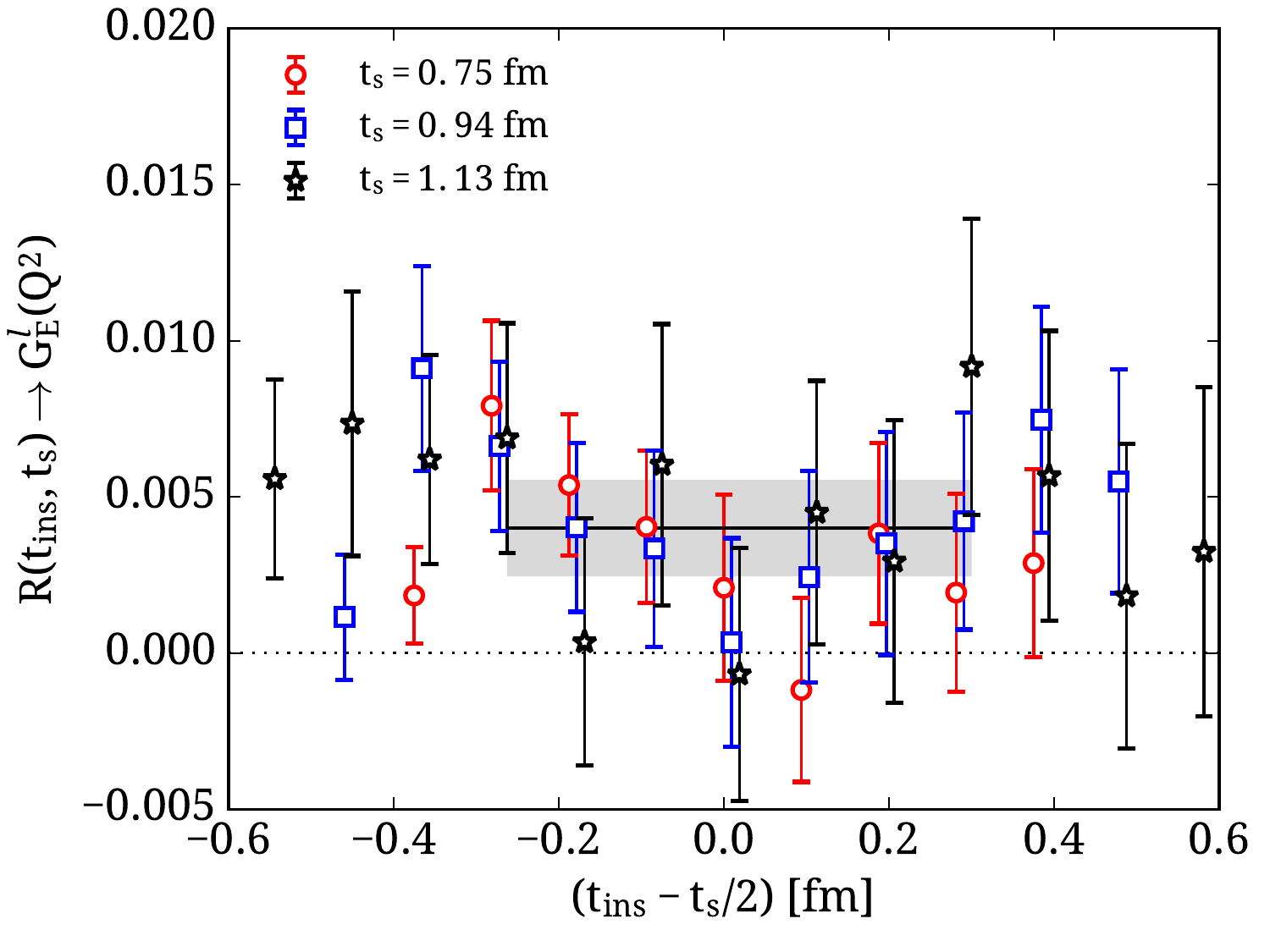}
  \caption{Results for the ratio from which $G_E^l(Q^2)$ is extracted. This is a representative example for $Q^2=0.0753$ GeV$^2$. The source-sink time separations are for $t_s=0.75$~fm (open red circles),  $t_s=0.94$~fm (open blue squares) and $t_s=1.13$~fm (open black stars). Results for the two larger separations are shifted slightly to the right for clarity. The gray band is the extracted value using the plateau method for $t_s=1.13$~fm, using $t_{\rm ins}$-values indicated by the length of the error band.}
  \label{fig:RGE}
\end{figure}

\begin{figure}[ht!]
  \includegraphics[width=\linewidth]{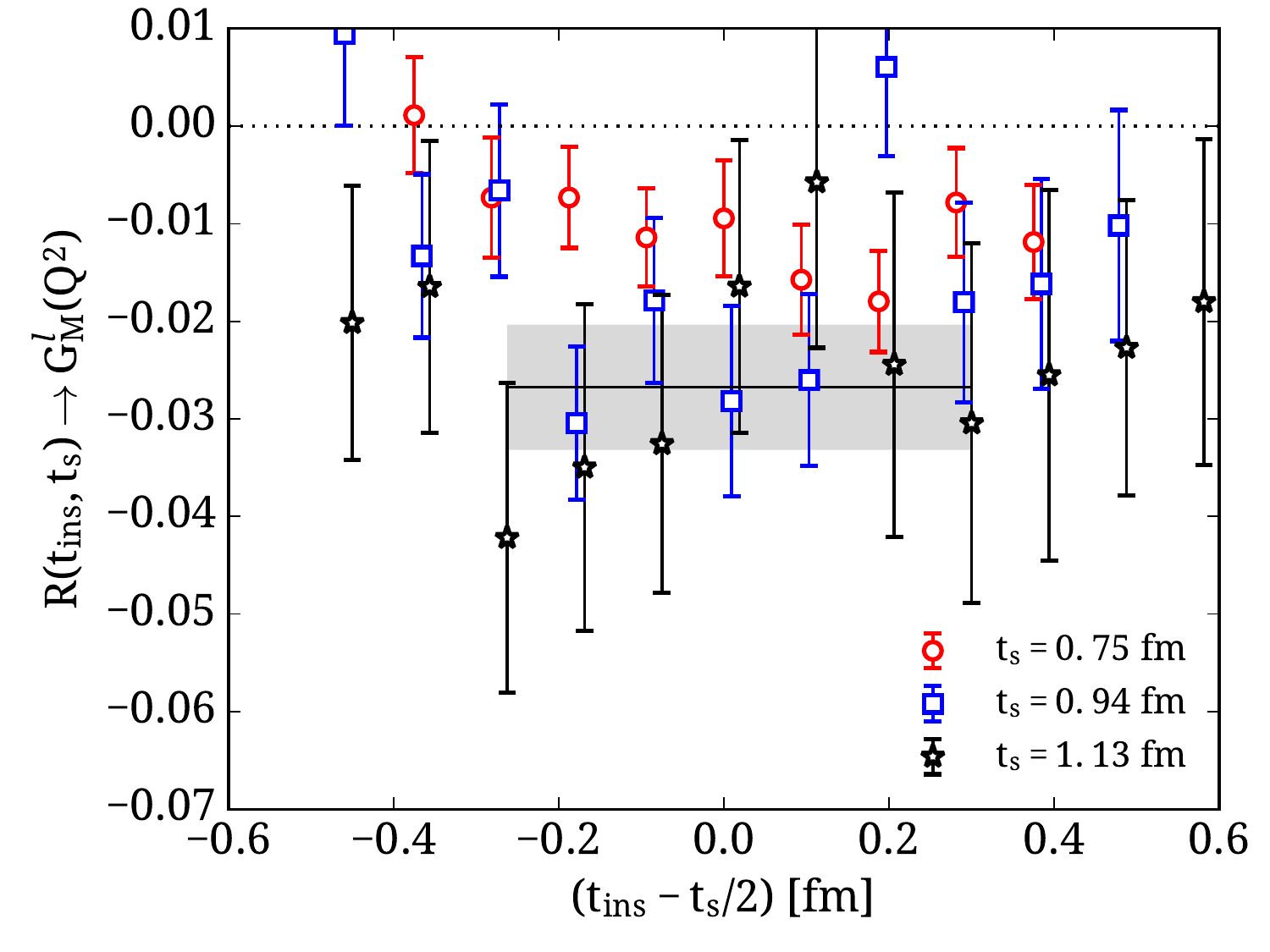}
  \caption{Ratio leading to $G_M^l(Q^2)$ for $Q^2=0.0753$ GeV$^2$. The notation is as in Fig.~\ref{fig:RGE}.}
  \label{fig:RGM}
\end{figure}

\begin{figure}[ht!]
  \includegraphics[width=\linewidth]{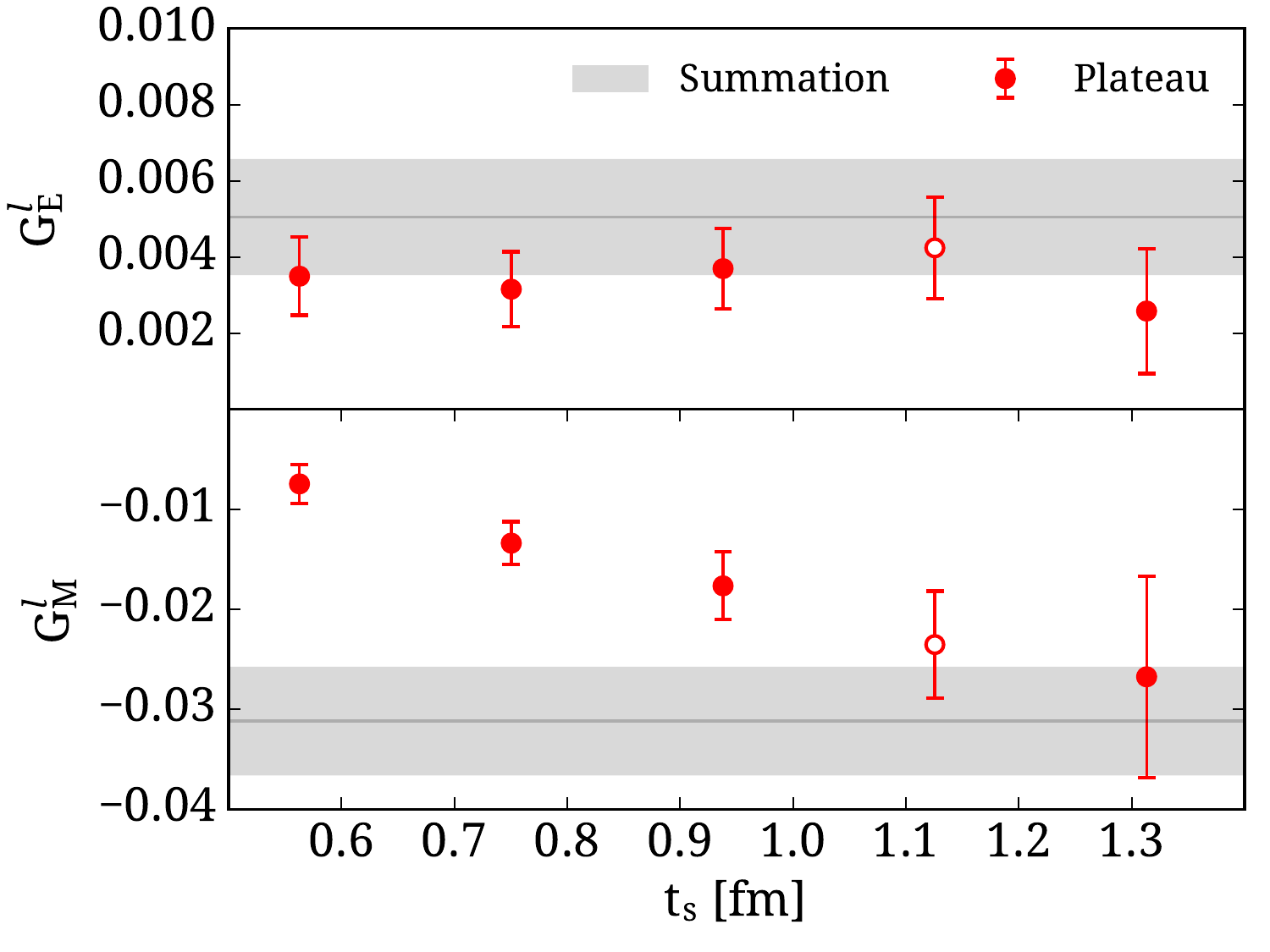}
  \caption{Extracted values for $G_E^l$ and $G_M^l$ at $Q^2=0.0753$ GeV$^2$ using the plateau method (red points) and summation method (gray band). Open symbols show our chosen value from the plateau method.}
  \label{fig:FF_vs_ts}
\end{figure}

We demonstrate the quality of our plateaus in Figs.~\ref{fig:RGE} and ~\ref{fig:RGM}.  The disconnected part of the  three-point function can be computed for all  source-sink time separations. However,  very large time separations are not useful due to the increased statistical error. Thus, we restrict  to analyzing separations up to $t_s=1.31$~fm for which the signal-to-noise ratio is acceptable.
In Fig.~\ref{fig:RGE} the ratio yielding $G_E^l(Q^2)$ is shown. Note that the upper index ``${\it l}$'' is used to denote the light quarks combination introduced in Eq.~(\ref{Eq:VecCurr_l}). For demonstration purposes we choose a representative momentum, namely $Q^2=0.0753$ GeV$^2$, having $\vec{p}\,'=\vec{0}$. 
In Fig.~\ref{fig:RGM} the ratio yielding $G_M^l(Q^2)$ is presented. Fitting the form factors within the plateau region for several separations allows us to check convergence to the ground state. The extracted  results are shown in Fig.~\ref{fig:FF_vs_ts} including also the result from the summation method obtained using the fit range [0.56-1.31]~fm. For the case of $G_E^l$, results using the plateau method up to $t_s=1.13$~fm have a good agreement with the summation method while larger separations become noisy. For $G_M^l$, the value increases in magnitude as $t_s$ increases and becomes compatible with the summation method for $t_s=1.13$~fm. Therefore, we show final results extracted using the plateau method at $t_s=1.13$~fm to which we perform our $Q^2$-fits in what follows. The same procedure is followed to extract the disconnected contributions to the form factors  at several $Q^2$ values where the analysis is extended to allow for non-zero final nucleon momentum yielding  a large number of closely spaced values for $Q^2$. To display the results  we do a weighted average on results with close values of $Q^2$. In particular, we use bins with width of $0.02$~GeV$^2$ for the light disconnected quark contributions and $0.04$~GeV$^2$ for the strange since  for the latter we have  results  available up to higher $Q^2$ compared to the light.  A systematic error due to excited states contamination is given by the difference between the plateau and the summation values. 


The dipole form is widely used to fit the proton electric and magnetic form factors~\cite{Hand:1963zz,Kelly:2004hm} yielding the
expected behavior in the large-$Q^2$ region where the form factors are expected to decrease like $Q^{-4}$~\cite{Perdrisat:2006hj}. The z-expansion~\cite{Hill:2010yb,Epstein:2014zua} is a model independent Ansatz that has been applied recently to fit experimental results.  Using a conformal mapping of $Q^2$ to a variable $z$ defined as,
\begin{equation}
z=\frac{\sqrt{t_{cut}+Q^2} - \sqrt{t_{cut}}}{\sqrt{t_{cut}+Q^2} + \sqrt{t_{cut}}}
\label{Eq:Zvar}
\end{equation}
one can expand the form factor into a polynomial
\begin{equation}
  G(Q^2) = \sum_{k=0}^{k_{max}} a_k z^k,
  \label{Eq:Zexp}
\end{equation}
where $t_{cut}$ is the cut in the time-like region of the form factor. For light disconnected form factors $t_{cut}=(2 m_\pi)^2$ is used while for the strange $t_{cut}=(2 m_K)^2$ with $m_K$ the kaon mass. The z-expansion should converge as we increase $k_{max}$ and the coefficients $a_k$ should be bounded in size for this to happen. The form factor at $Q^2=0$ is obtained from the first coefficient, i.e. $G(Q^2=0)=a_0$. We define the radius as,
\begin{equation}
   r^2  = -6 \frac{d G(Q^2)}{dQ^2} \Big \vert_{Q^2=0},
  \label{Eq:Radius}
\end{equation}
which is related to the second coefficient, via $r^2 = -3 a_1/2 t_{cut}$. In the case of the proton and neutron electric form factors the mean square radius is the same as Eq.~(\ref{Eq:Radius}), whereas for the magnetic, one has to divide with the total value of the form factor at $Q^2=0$.

In our fitting procedure, the coefficients $a_0,\,a_1$ are free to vary, while for $a_{k>1}$ we impose Gaussian priors for the series to converge. The priors are imposed using an augmented $\chi^2$ where the additional term is
\begin{equation}
  \chi^2_{pr} = \sum_{k>1}^{k_{max}} \frac{(a_k-\tilde{a}_k)^2}{w_{a_k}}
\end{equation}
for parameter $a_k$, which is centered at $\tilde{a}_k$ with width $w_{a_k}$. To compute $\tilde{a}_k$ we start by setting $k_{max}=1$ to obtain an estimate for $a_0$ and $a_1$ using jackknife ensemble averages. Then, for $k_{max}=2$, $\tilde{a}_2$ is set to ${\rm max(\vert a_0 \vert, \vert a_1 \vert)}$ and the width is chosen as $w_{a_k}=2\, \vert \tilde{a}_k \vert$.
This procedure is generalized for any $k_{max}$ and the priors are used to restrict $a_k$ inside the jackknife bins.
\begin{figure}[ht!]
  \includegraphics[width=\linewidth]{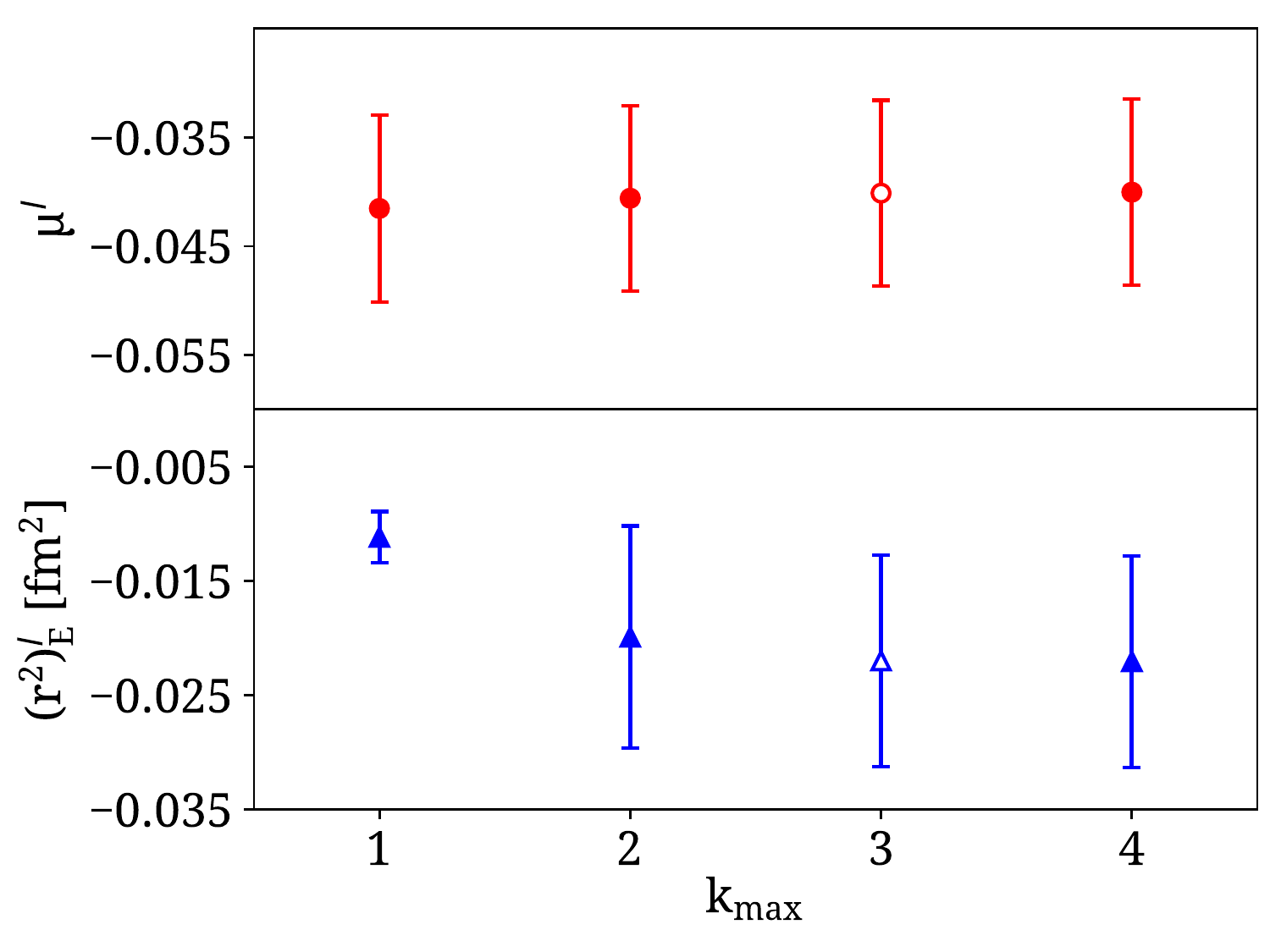}
  \caption{Extracted values for $\mu^l$ and $( r^2 )_E^l$ as a function of $k_{\rm max}$, where results from the plateau method at $t_s=1.13$~fm have been used.}
  \label{fig:Vals_vs_Kmax}
\end{figure}
In Fig.~\ref{fig:Vals_vs_Kmax} we show two representative observables extracted from the electromagnetic form factors using the z-expansion as a function of $k_{\rm max}$. We seek for convergence in both mean value and error as we increase $k_{\rm max}$. In the case of the magnetic moment $\mu^l$, increasing $k_{\rm max }$ does not affect the result while in the case of the radius $( r^2 )_E^l$ one needs up to $k_{\rm max}=3$ to converge. Therefore, we choose to use $k_{\rm max}=3$ for all the extracted quantities where we have checked the convergence of Eq.~(\ref{Eq:Zexp}).

\begin{figure}[ht!]
  \includegraphics[width=\linewidth]{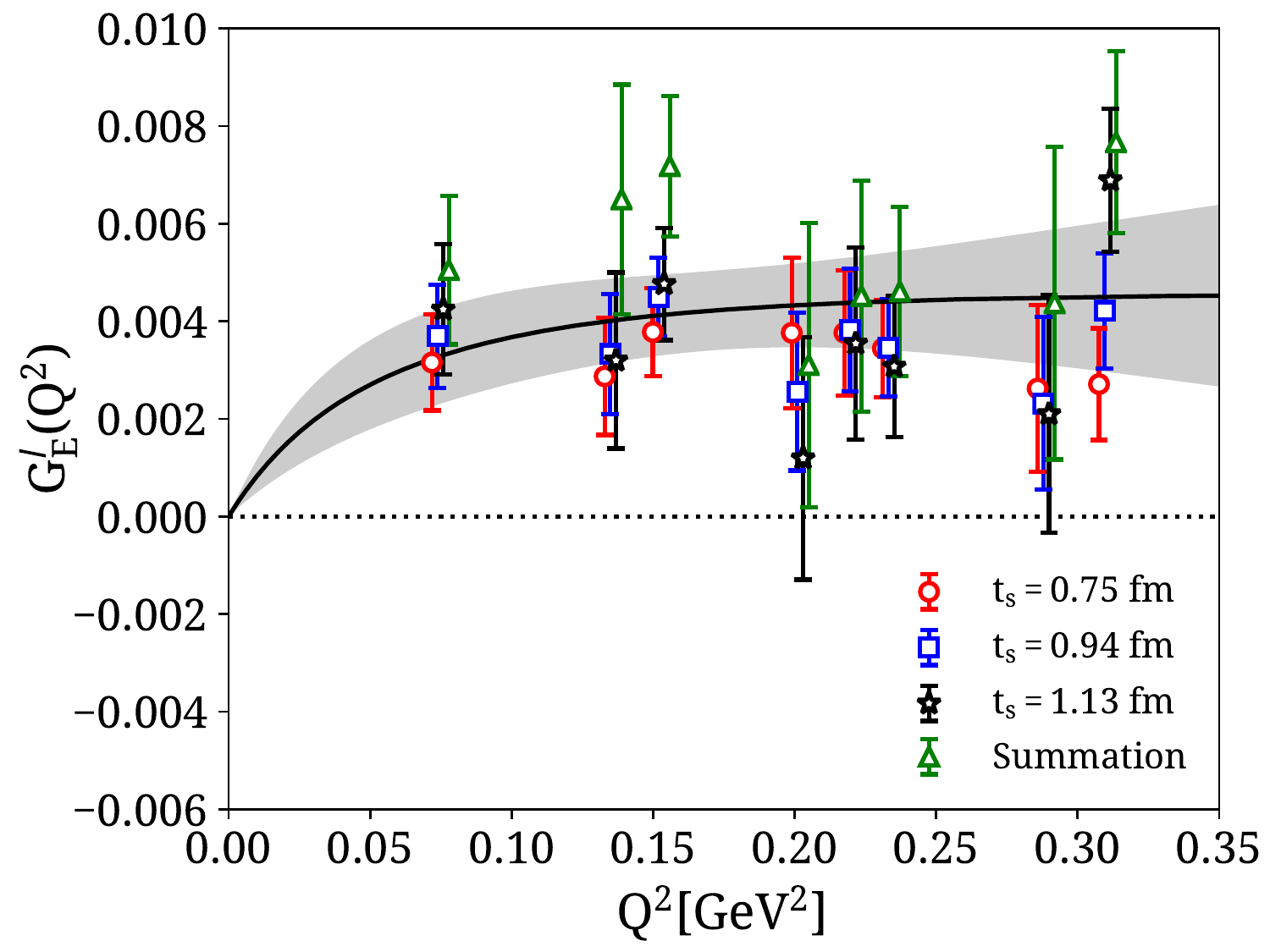}
  \caption{Disconnected light quarks contribution to the nucleon electric form factor denoted as $G_E^l(Q^2)$. Results are extracted using the plateau method for three source-sink time separations  with $t_s=0.75$~fm (red open circles), $t_s=0.94$~fm (blue open squares) and $t_s=1.13$~fm (black open stars). Results using the summation method in the fit range of [0.56-1.31]~fm are depicted with the green open triangles. Results shown are obtained after a binning of neighboring $Q^2$ values as explained in the text. Results are shifted slightly to the right for clarity. The gray band is a fit to the results extracted from the plateau method using $t_s=1.13$~fm.}
  \label{fig:GE_l}
\end{figure}

In Fig.~\ref{fig:GE_l} we present the light quarks disconnected contribution to the nucleon electric form factor. The form factor is shown up to $Q^2 \sim 0.3$~GeV$^2$. The fits of the form factor yield a monotonically increasing dependence on the $Q^2$ that flattens out for $Q^2>0.2$~GeV$^2$.
In the case of $G_E^l(Q^2)$ we impose  $a_0=0$. Fitting the results extracted from the plateau method at $t_s=1.13$~fm, we find a value for the radius $( r^2 )_E^l = -0.022(9)$ $\rm fm^2$,
whereas using the summation method we find $( r^2 )_E^l = -0.035(11)$ $\rm fm^2$.
We assign a systematic error due to possible excited states from
the difference between the values extracted using the plateau and summation methods obtaining a value for the electric squared charge radius 
\begin{equation}
  ( r^2 )_E^l = -0.022(9)(13)~\textrm{fm}^2.
\end{equation}

It is interesting to check how much the proton and neutron charge radii are affected by the disconnected contributions. Using results for the connected contributions from Ref.~\cite{Alexandrou:2017ypw}, tabulated in Table~\ref{table:Results}, we find that the connected plus disconnected light quark contributions are
\begin{eqnarray}
  ( r^2 )_E^p (\textrm{total}) &=& 0.562(31)(31)~\textrm{fm}^2,\\
  ( r^2 )_E^n (\textrm{total}) &=& -0.064(25)(14)~\textrm{fm}^2. 
\end{eqnarray}
Although the light disconnected contribution to the proton charge radius is small, it is important to calculate accurately enough when comparing to experiment, especially in light of the discrepancy observed in the experimental value of proton charge radius between the conventional and the muonic hydrogen measurement. For the neutron, disconnected quark contributions are more important making the value of the charge radius more negative, albeit with large statistical errors.

\begin{figure}[ht!]
  \includegraphics[width=\linewidth]{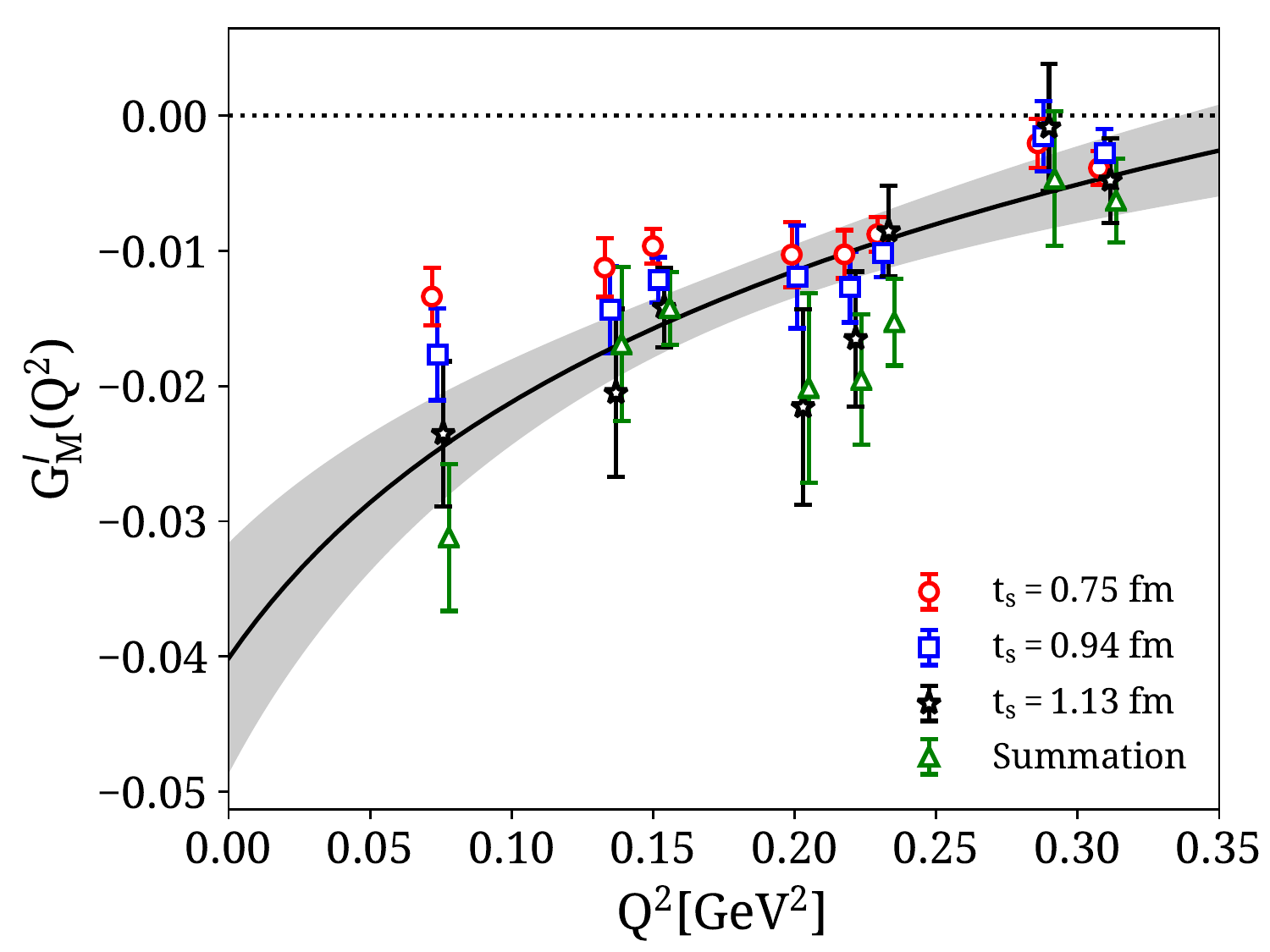}
  \caption{Disconnected light quarks contribution to the nucleon magnetic form factor $G_M^{l}(Q^2)$. The notation is as in Fig.~\ref{fig:GE_l}.}
  \label{fig:GM_l}
\end{figure}

\begin{figure}[ht!]
  \includegraphics[width=\linewidth]{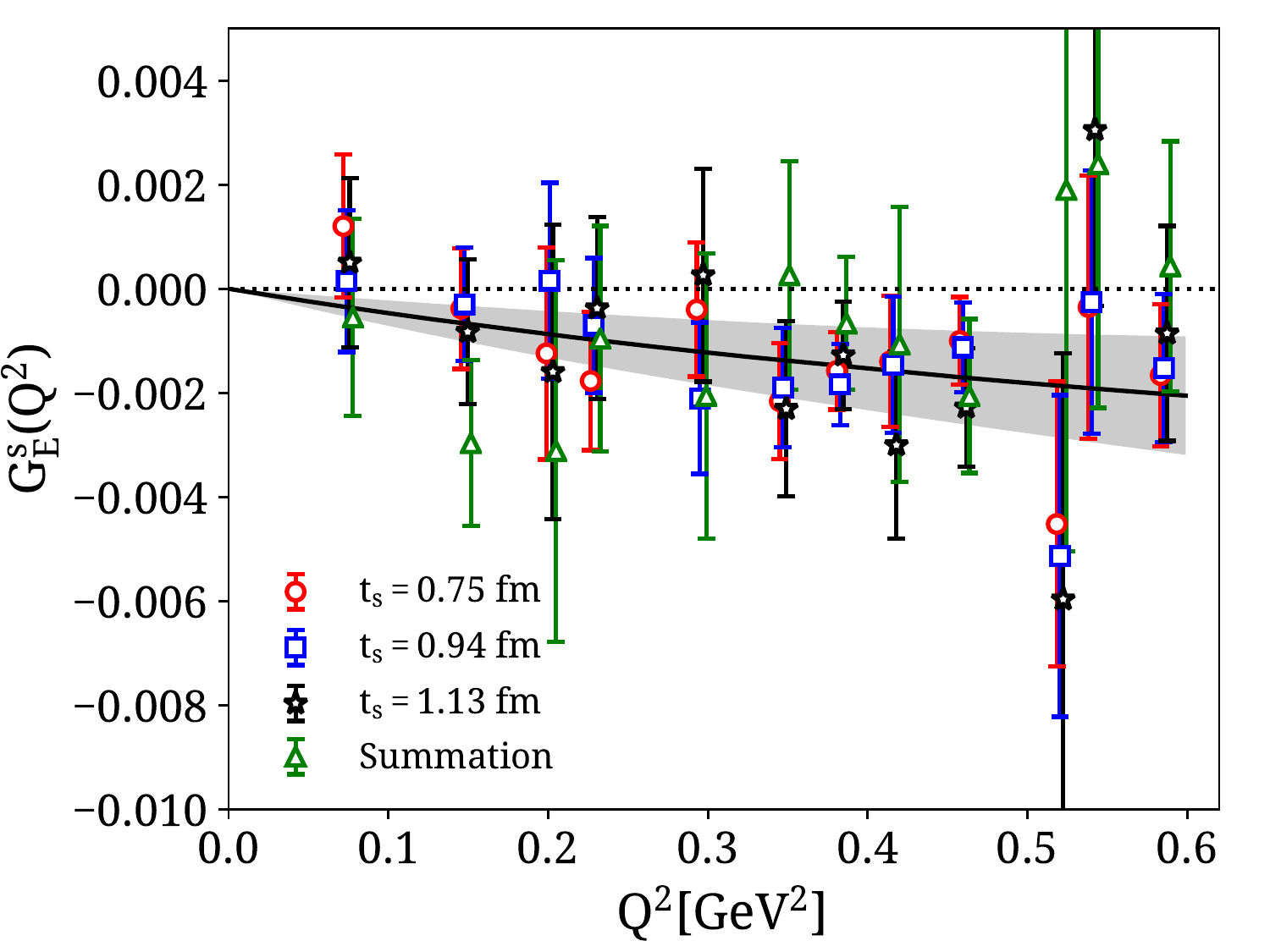}
  \caption{Strange  nucleon electric form factor, $G_E^s(Q^2)$. The notation is as in Fig.~\ref{fig:GE_l}.}
  \label{fig:GE_s}
\end{figure}

In Fig.~\ref{fig:GM_l} we show our results for $G_M^{l}(Q^2)$, which as noted above, shows a clear trend to decrease by increasing source-sink time separation, especially at small values of $Q^2$.
Fitting $G_M^{l}(Q^2)$ using the z-expansion we find that disconnected contributions to the nucleon magnetic moment and radius are $\mu^l = -0.040(9)(3), \;\;\; ( r^2 )^{l}_M = -0.071(24)(4)~\textrm{fm}^2$.
In Fig.~\ref{fig:GE_s}, we show results for  the strange nucleon electric form factor, which  receives only disconnected contributions.
We find that the strange charge radius of the nucleon is 
\begin{equation}
  ( r^2 )^s_E = 0.0012(6)(7)~\textrm{fm}^2,
\end{equation}
which is consistent with zero if one takes into account the systematic error due to the estimate of excited state contributions. 

\begin{figure}[ht!]
  \includegraphics[width=\linewidth]{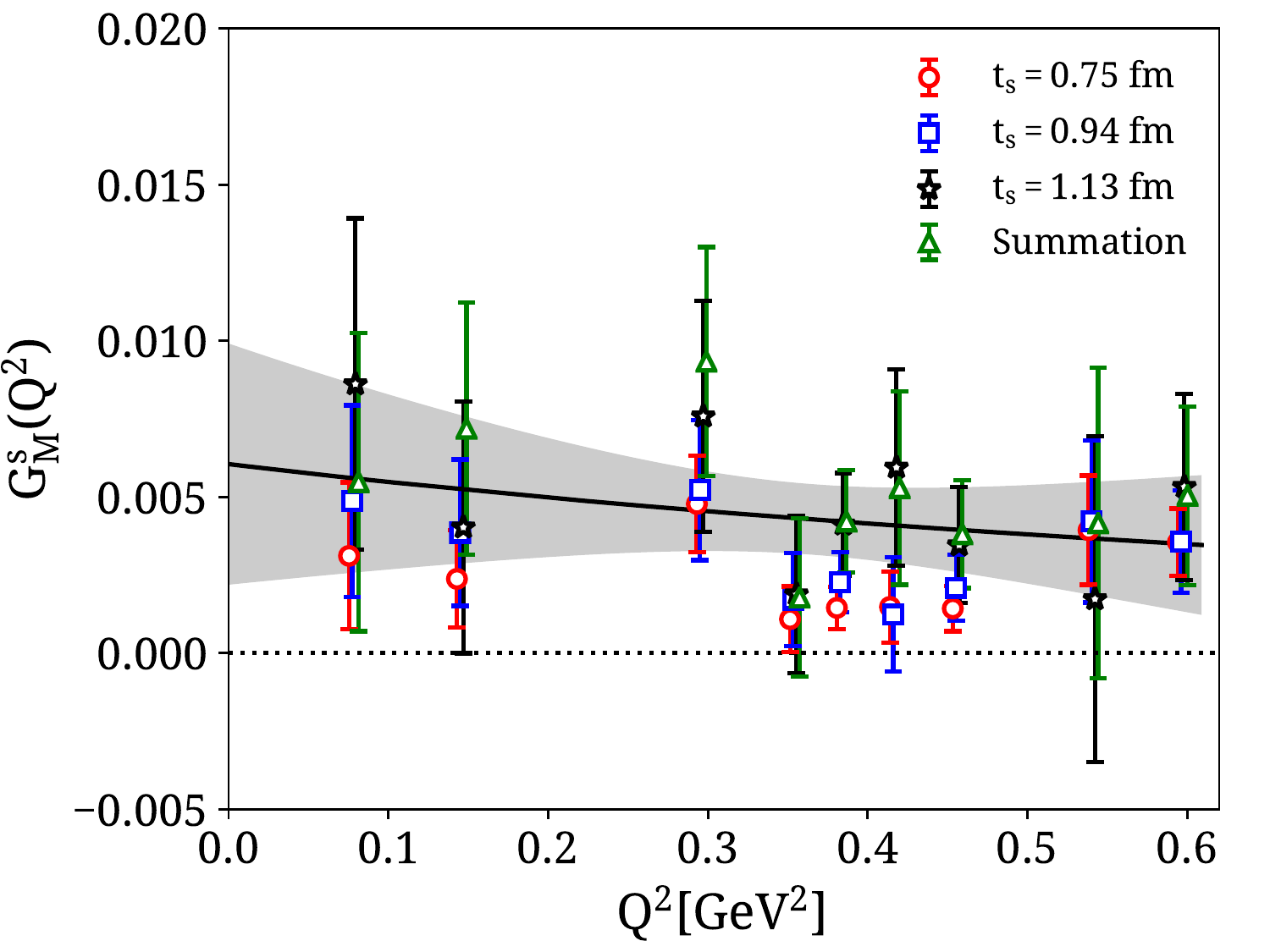}
  \caption{Strange  nucleon magnetic form factor, $G_M^s(Q^2)$. The notation is as in Fig.~\ref{fig:GE_l}.}
  \label{fig:GM_s}
\end{figure}

The strange magnetic form factor $G_M^s(Q^2)$ is shown in Fig.~\ref{fig:GM_s}.
We find a strange nucleon  magnetic moment of
\begin{equation}
  \mu^s = 0.006(4)(1).
\end{equation}
The strange magnetic radius is $( r^2 )^s_M = 0.0014(27)(2)~\textrm{fm}^2$, consistent with zero, as expected from the flat behavior of the form factor in Fig.~\ref{fig:GM_s}. Our results for the proton and neutron magnetic moments and radii are  given in Table~\ref{table:Results}.

\setlength{\tabcolsep}{0.5em}
\begin{widetext}
  \begin{center}
    \begin{table}[ht!]
  \caption{Our final results for $( r^2 )_E$ (first row), $\mu$ (middle row) and $( r^2 )_M$ (last row). In the first and second columns we give the light and strange disconnected contributions, in the third and fourth, the proton connected and total values and in the fifth and sixth the corresponding ones for  neutron. The radius is defined in Eq.~(\ref{Eq:Radius}). In the case of the magnetic radius one has to divide with the total value of magnetic moment to extract the mean square radius. Results for the connected are taken from Ref.~\cite{Alexandrou:2017ypw}.}
  \label{table:Results}      
  \begin{tabular}{c | c | c || c | c | c | c }
    \hline
    Quantity & Disc. light & Strange & p (conn.) & p (total) & n (conn.) & n (total)\\
    \hline \hline
    $( r^2 )_E$ [fm$^2$] & -0.022(9)(13) & 0.0012(6)(7) & 0.584(30)(28) & 0.563(31)(31) & -0.042(23)(6) & -0.063(25)(14) \\
    $\mu$ &-0.040(9)(3) & 0.006(4)(1) & 2.455(127)(155) & 2.421(127)(155) & -1.564(94)(123) & -1.598(95)(123) \\
    $( r^2 )_M$ [fm$^2$] & -0.071(24)(4) & 0.0014(27)(2) & 1.284(183)(218) & 1.214(185)(218) & -0.875(139)(180) & -0.945(141)(180) \\
    \hline
  \end{tabular}
\end{table}
\end{center}
\end{widetext}

\section{Comparison with other studies} \label{Sec:Comp}
Disconnected quark loop contributions to the nucleon electromagnetic form factors are available from two recent works beyond the current one. In Ref.~\cite{Green:2015wqa}, LHPC has analyzed an ensemble of  $N_f=2+1$ Wilson clover-improved fermions simulated for heavier than physical pion mass, namely $m_\pi=317$~MeV.
The other study, from $\chi$QCD, used valence overlap fermions on four $N_f=2+1$ domain-wall fermion ensembles with pion masses in the range $m_\pi \in (135,403)$~MeV~\cite{Sufian:2017osl}. Their final values were extracted by performing a simultaneous chiral, infinite volume and continuum extrapolation.

\begin{figure}[ht!]
  \includegraphics[width=\linewidth]{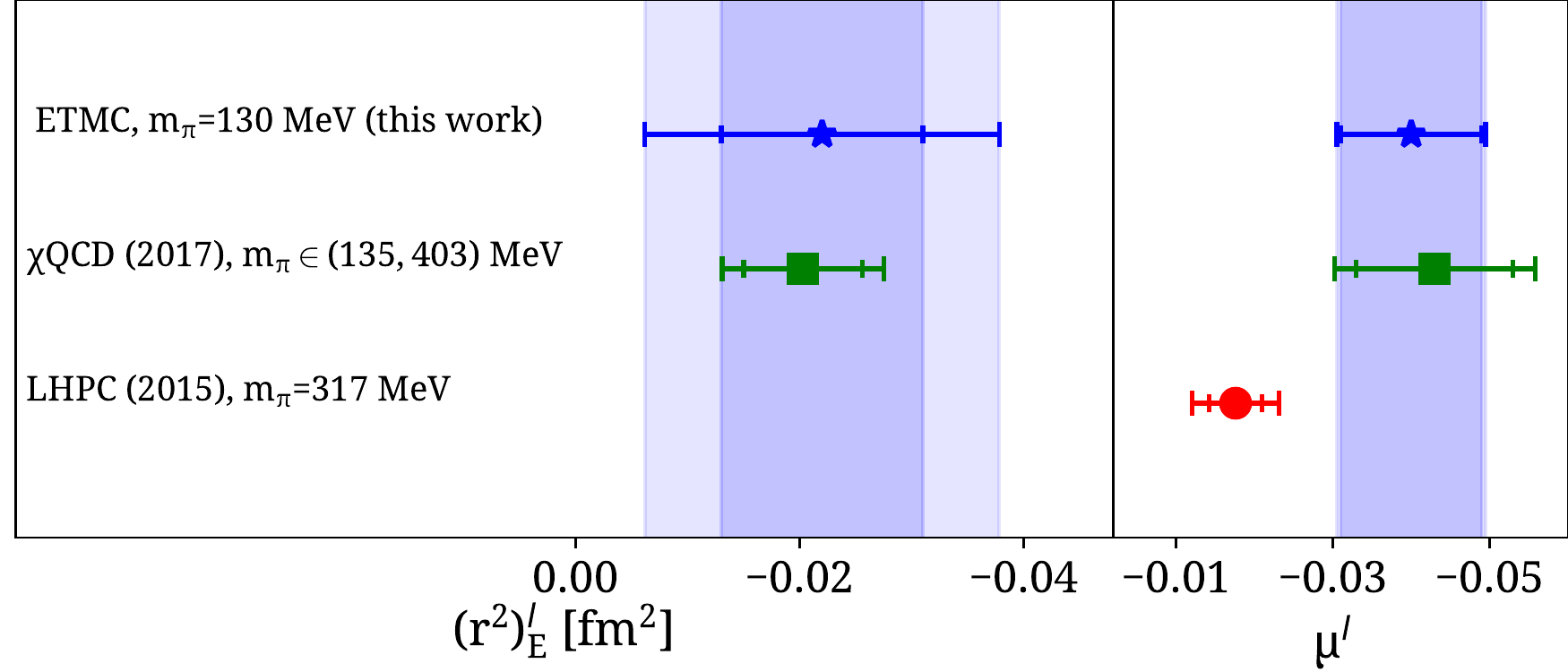}
  \caption{Comparison of our results (blue star) for $\mu^l$ with results from LHPC (red circle) and $\chi$QCD (green square) and for $( r^2 )^{l}_E$ with $\chi$QCD. We multiply by a factor of 1/3 the results from LHPC to match our convention. The inner error band is the statistical error, while the outer band is the total error. }
  \label{fig:CompLight}
\end{figure}

\begin{figure}[ht!]
  \includegraphics[width=\linewidth]{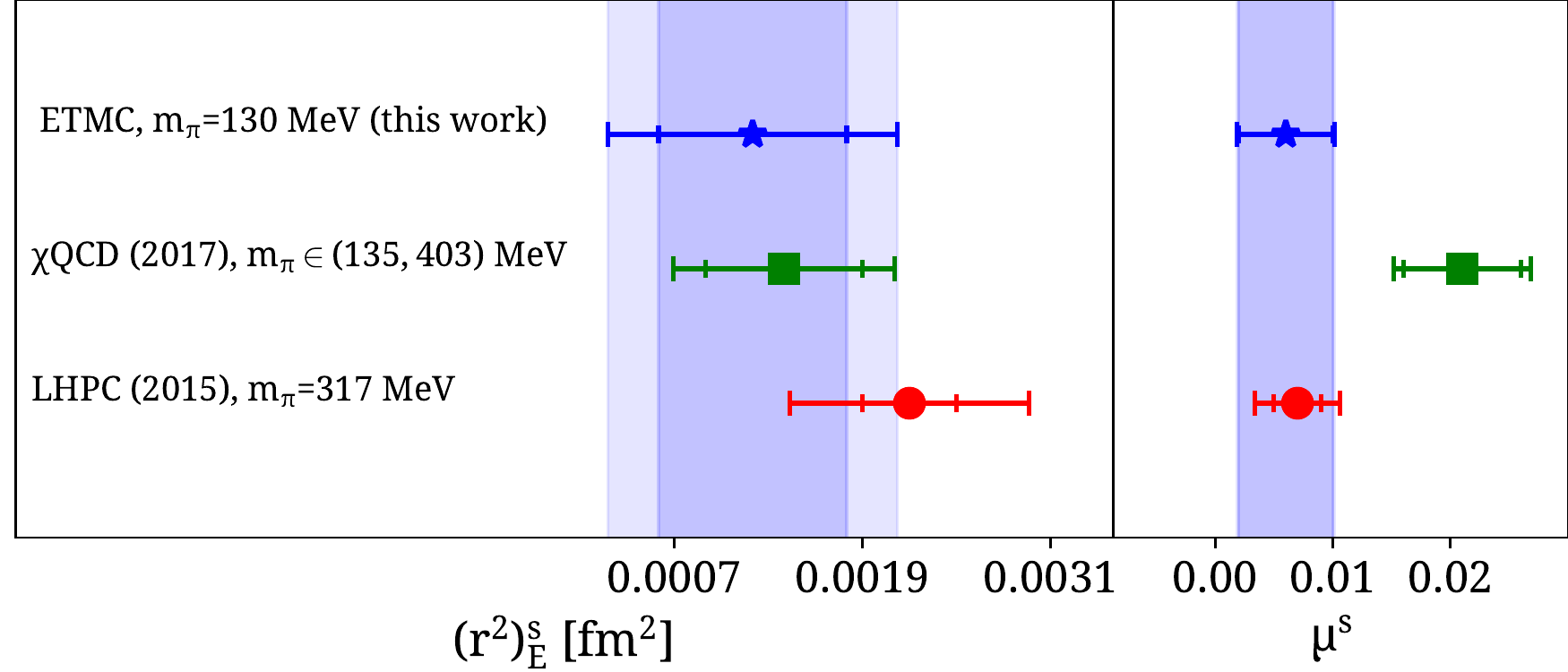}
  \caption{Comparison of our results (blue star) for $( r^2 )^s_E$ and  $\mu^s$ with results from LHPC (red circles) and $\chi$QCD (green square). The convention is as in Fig.~\ref{fig:CompLight}. }
  \label{fig:CompStrange}
\end{figure}

In Fig.~\ref{fig:CompLight}, we compare our result for $\mu^{l}$ to the one from $\chi$QCD, while for $( r^2 )^{l}_E$ to those from both $\chi$QCD and LHPC. The dark, inner band indicates the statistical error, while the outer band is the statistical and systematic error added in quadrature. The good agreement with $\chi$QCD, for which a continuum and infinite volume extrapolation has been performed, indicates that lattice artifacts due to finite lattice spacing and volume on these quantities are small for our ensemble. On the other hand, the result for $\mu^l$ from LHPC at higher than physical pion mass is smaller, as expected from chiral perturbation theory arguments. In Fig.~\ref{fig:CompStrange} we compare the strange  $\mu^s$ and $( r^2 )^s_E$ with the corresponding results from the two other studies. For $( r^2 )^s_E$, results from the three studies are in good agreement, whereas for $\mu^s$, the result from $\chi$QCD agrees within one standard deviation. Given the large statistical errors on the strange quark contributions such an agreement among lattice QCD results is welcoming.

\section{Conclusions} \label{Sec:Concl}
In this study, we compute the disconnected quark loop contributions from up, down and strange quarks to the nucleon electromagnetic form factors using $N_f=2$ maximally twisted mass fermions at the physical point. While all source-sink time separations accessible, we opt to use up to $t_s=1.31$~fm for which statistical errors are not prohibitively large. Both the plateau and the summation methods are employed to estimate contamination due to the excited states. Three-point functions produced with final nucleon momenta of $\vec{p}\,'=\vec{0}$ and $\vec{p}\,'=\frac{2 \pi}{L} \vec{\hat{n}}$ and analyzed to increase statistics. The form factors, $G_E^l(Q^2)$ and $G_M^l(Q^2)$, are computed up to $Q^2 \simeq 0.3$~GeV$^2$ while $G_E^s(Q^2)$ and $G_M^s(Q^2)$ are computed up to $Q^2 \simeq 0.6$~GeV$^2$. The model independent z-expansion is used to fit the $Q^2$ dependence of the form factors and  extract the electric and magnetic radii as well as the magnetic moment. The size of the individual contributions as well as the total values for the extracted quantities are tabulated in Table~\ref{table:Results}. While the contribution of the light quark-disconnected diagram is clearly non-zero, strange quark contributions are almost consistent with zero within the current errors.

We plan to analyze an $N_f=2+1+1$ twisted mass ensemble with a clover term at the physical point to check possible quenching effects of the strange and charm quarks in the sea. Further improvements for the computation of  disconnected quark loops are under investigation to improve the accuracy of the disconnected loop determination.

\textit{Acknowledgments:} We would like to thank the members of the
ETM Collaboration for a productive collaboration. We acknowledge
funding from the European Union's Horizon 2020 research and innovation
program under the Marie Sklodowska-Curie Grant Agreement
No. 642069.We gratefully acknowledge the Gauss Centre for
Supercomputing e.V. (\url{www.gauss-centre.eu}) for funding this
project by providing computing time on the GCS Supercomputer SuperMUC
at Leibniz Supercomputing Centre (\url{www.lrz.de}).
Results were
obtained using Piz Daint at Centro Svizzero di Calcolo Scientifico
(CSCS), via projects with ids s540, s625 and s702.  We thank the staff
of CSCS for access to the computational resources and for their
constant support as well as the J\"ulich Supercomputing Centre (JSC)
for the tape storage. MC acknowledged financial support by the US
National Science Foundation under Grant No. PHY-1714407.
\clearpage
\bibliographystyle{unsrtnat}
\bibliography{refs.bib}

\appendix
\widetext
\section{Extraction of form factors from lattice QCD ratios}
\label{sec:appendix equations}

In this Appendix we generalize the equations from which the form factors are extracted for a  nucleon with non-zero final momentum ${\vec p}^\prime$. All expressions are given in Euclidean space.
\begin{eqnarray}
\Pi_\mu(\Gamma_0,\vec{p}\,',\vec{p})) &=& \frac{-i G_E(Q^2) C}{2 m (4 m^2 + Q^2)} \left( (p^\prime_\mu + p_\mu) \left[m\left(E(\vec{p^\prime}) + E(\vec{p}) +m\right) -p^\prime_\rho p_\rho\right] \right) \nonumber \\
&&   + \frac{G_M(Q^2) C}{4 m^2(4m^2+Q^2)} \Big( \delta_{\mu 0} \big( 4m^4 + m^2Q^2 + 4m^2 p^\prime_\rho p_\rho + Q^2 p^\prime_\rho p_\rho \big)  \nonumber \\
&&   + 2im^2p^\prime_\mu \big(E(\vec{p^\prime}) - E(\vec{p})\big) - 2im^3(p^\prime_\mu+p_\mu) - E(\vec{p})iQ^2p^\prime_\mu - E(\vec{p^\prime})iQ^2p_\mu \nonumber \\
&&  - imQ^2 (p^\prime_\mu+p_\mu) - 2im^2p_\mu \big(E(\vec{p^\prime}) - E(\vec{p})\big) - 2 i m p^\prime_\rho p_\rho (p^\prime_\mu+p_\mu) \Big);
\label{Eq:GEGMEQ1}
\end{eqnarray}

\begin{eqnarray}
\Pi_\mu(\Gamma_k,\vec{p}\,',\vec{p})) &=&  \frac{-G_E(Q^2) C}{2 m (4 m^2 + Q^2)} \Big( m^2 \varepsilon_{\mu k 0 \rho} (p_\rho^\prime - p_\rho) - i \varepsilon_{\mu k \rho \sigma} p_\rho^\prime p_\sigma \big(E(\vec{p^\prime}) + E(\vec{p})\big) \nonumber \\
&& + \varepsilon_{\mu 0 \rho \sigma}p_\rho^\prime p_\sigma (p_k^\prime + p_k) - \varepsilon_{\mu k 0 \rho} p_\sigma^\prime p_\sigma (p_\rho^\prime - p_\rho) \Big) \nonumber \\
&&  - \frac{G_M(Q^2) C}{4 m^2(4m^2+Q^2)} \Big( m \varepsilon_{\mu k 0 \rho} (p_\rho^\prime - p_\rho) (2m^2 + Q^2)  \nonumber \\
&&  + 2im \varepsilon_{\mu k \rho \sigma} p^\prime_\rho p_\sigma \big(2m+E(\vec{p^\prime}) + E(\vec{p})+\frac{Q^2}{2m} \big) \nonumber \\
&&  - 2m\varepsilon_{\mu 0 \rho \sigma}p^\prime_\rho p_\sigma (p_k^\prime + p_k) + 2m\varepsilon_{\mu k 0 \rho} p^\prime_\sigma p_\sigma (p_\rho^\prime - p_\rho) \Big)
\label{Eq:GEGMEQ2},
\end{eqnarray}

where 
\begin{equation}
  C= \frac{2m}{E(\vec{p})(E(\vec{p}\,')+m)} \sqrt{\frac{E(\vec{p})(E(\vec{p}\,')+m)}{E(\vec{p}\,')(E(\vec{p})+m)}}.
\end{equation}

\end{document}